\documentclass[10pt,twocolumn,letterpaper]{article}

\usepackage{iccv}
\usepackage{authblk}
\usepackage{times}
\usepackage{epsfig}
\usepackage{graphicx}
\usepackage{amsmath}
\usepackage{amssymb}
\usepackage{enumitem}
\usepackage{multirow}
\usepackage{booktabs}

\usepackage[T1]{fontenc}
\usepackage[utf8]{inputenc}
\usepackage[english]{babel}
\usepackage[autostyle, english=american]{csquotes}
\MakeOuterQuote{"}

\usepackage[breaklinks=true,bookmarks=false]{hyperref}

\iccvfinalcopy 

\def\iccvPaperID{****} 

\ificcvfinal\fi

\begin{document}
\title{The Solution for Temporal Sound Localisation Task of ICCV 1st Perception Test Challenge 2023}

\author{
Yurui Huang $^1$,
Yang Yang $^1$\thanks{Corresponding author: Yang Yang (yyang@njust.edu.cn)}, 
Shou Chen $^2$,
Xiangyu Wu $^{1,3}$,
Qingguo Chen $^3$,
Jianfeng Lu $^1$
}

\affil{
$^1$Nanjing University of Science and Technology\\
$^2$RIKEN Center for Advanced Intelligence Project, Japan\\
$^3$Alibaba
}

\maketitle
\ificcvfinal\thispagestyle{empty}\fi

\begin{abstract}
In this paper, we propose a solution for improving the quality of temporal sound localization. We employ a multimodal fusion approach to combine visual and audio features. High-quality visual features are extracted using a state-of-the-art self-supervised pre-training network, resulting in efficient video feature representations. At the same time, audio features serve as complementary information to help the model better localize the start and end of sounds. The fused features are trained in a multi-scale Transformer for training. In the final test dataset, we achieved a mean average precision (mAP) of 0.33, obtaining the second-best performance in this track.
\end{abstract}

\section{Introduction}

Temporal sound localization is the task of determining the start time and end time of sound events in a complete video. A similar task to this is temporal action localization. In most previous works, action proposals\cite{DBLP:conf/iccv/LinLLDW19,yang2019comprehensive} or anchor windows \cite{DBLP:conf/cvpr/LongYQTLM19,yang2019semi} have been considered, and convolutional \cite{DBLP:conf/cvpr/ShouCZMC17,DBLP:journals/ijcv/ZhaoXWWTL20,yy2022}, recurrent\cite{DBLP:conf/bmvc/BuchEGFN17,yang2018complex}, and graph  neural networks \cite{DBLP:conf/aaai/YangQ022,DBLP:conf/cvpr/XuZRTG20, DBLP:conf/eccv/BaiWTYLL20,yy2023} have been developed for Temporal Action Localization (TAL). While stable progress has been made on major benchmark tests, the accuracy of existing methods often comes at the cost of modeling complexity. This complexity arises from increasingly complex proposal generation, anchor design, loss functions, network architectures, and output decoding processes.

Taking inspiration from the recent success of Transformers in the fields of NLP and computer vision, Transformers have also been applied to TAL tasks. For example, \cite{DBLP:conf/eccv/ZhangWL22,yang2023towards} combines multi-scale feature representations with local self-attention, and uses lightweight decoders for moment-level classification, and estimates corresponding action boundaries. \cite{DBLP:conf/cvpr/ShiZCMLT23,} introduces a novel trident head that models action boundaries by estimating the relative probability distribution around the boundaries. It also proposes an effective Scalable Granularity Perception (SGP) layer to mitigate the rank loss problem caused by self-attention in video features and aggregates information across different time granularities.

In this paper, we use a Transformer-based method called Actionformer as the backbone network. To obtain better feature representation, we employ the large-scale pre-trained network VideoMAE V2 \cite{DBLP:conf/cvpr/WangHZTHWWQ23} for visual encoding of videos. We also integrate multimodal information by fusing the visual and audio modalities at the early fusion stage. As a result, we achieved the second-best performance in the Sound Localization (SL) task with a mean Average Precision (mAP) of 0.33.

\section{Related Work}

\textbf{Self-supervised pre-training}: Due to the difficulty of labeled data, pre-training on large-scale pre-training datasets and fine-tuning on target datasets has become a mainstream training paradigm in many domains. For example, in NLP, Bert \cite{DBLP:conf/naacl/DevlinCLT19} is pre-trained on a large amount of text using Masked Language Modeling (MLM) and Next Sentence Prediction (NSP), becoming a widely used network in NLP. MAE \cite{DBLP:conf/cvpr/HeCXLDG22} masks random patches of input images and reconstructs the missing pixels, outperforming supervised learning. In the field of video understanding, Wang  \cite{DBLP:conf/cvpr/WangHZTHWWQ23}introduces the idea of MAE into self-supervised masked training of videos, achieving good results in various domains.

\textbf{Temporal action/sound localization}: Most previous works, action proposals\cite{DBLP:conf/iccv/LinLLDW19} or anchor windows \cite{DBLP:conf/cvpr/LongYQTLM19} have been considered, and convolutional \cite{DBLP:conf/cvpr/ShouCZMC17,DBLP:journals/ijcv/ZhaoXWWTL20}, recurrent\cite{DBLP:conf/bmvc/BuchEGFN17}, and graph neural networks \cite{DBLP:conf/aaai/YangQ022,DBLP:conf/cvpr/XuZRTG20, DBLP:conf/eccv/BaiWTYLL20} have been developed for Temporal Action Localization (TAL). With the development of Transformers, some works have applied them to the TAL domain. For example, Girdhar \cite{DBLP:conf/cvpr/GirdharCDZ19} use 2D object proposals as input to address the problem of temporal action localization. The problem of temporal sound localization can often be approached using methods from temporal action localization.
\begin{figure}
\begin{center}
\includegraphics[width=0.5\textwidth]{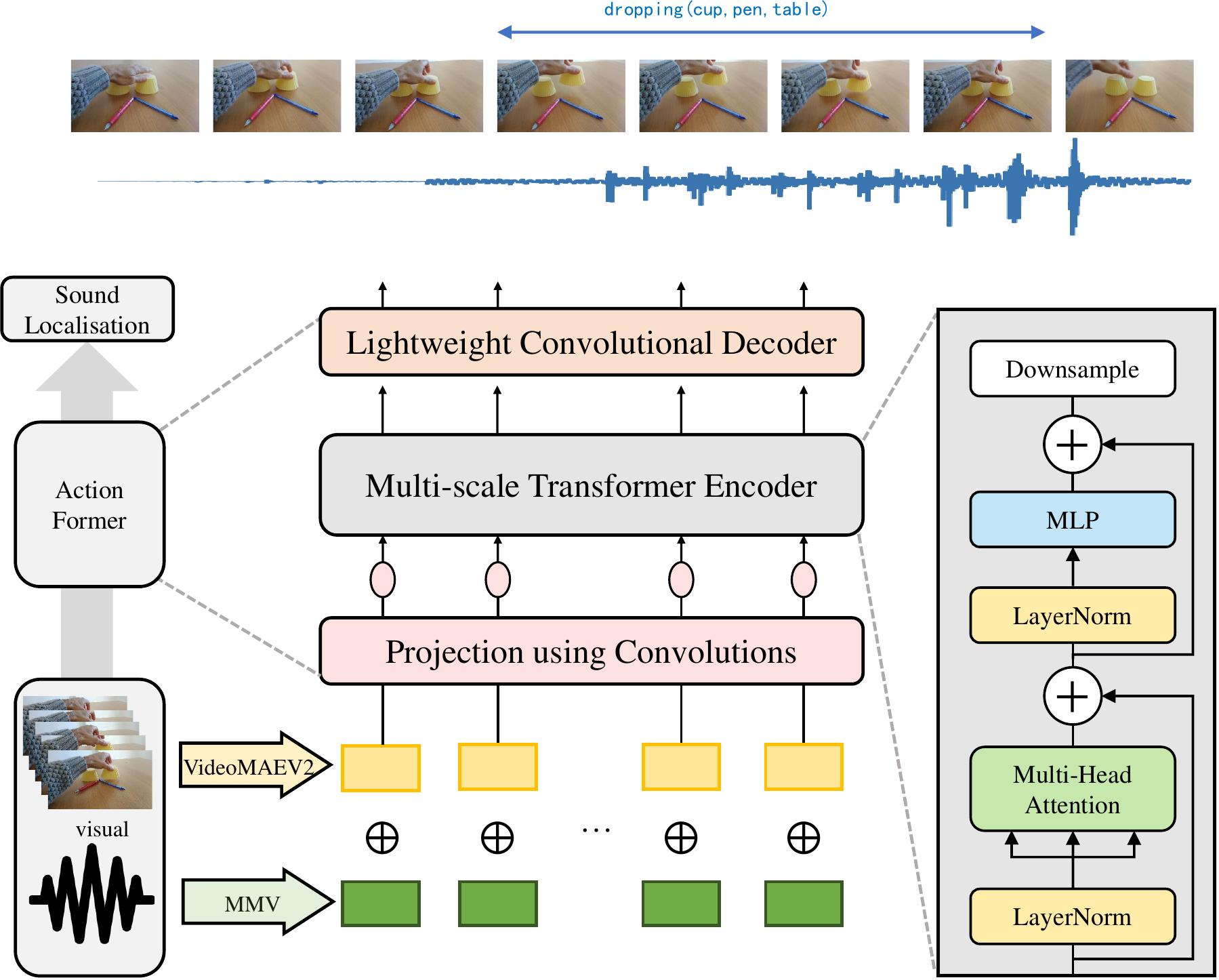}
\end{center}
   \caption{The visual and audio modalities are inputted, with the visual modality using VideoMAE V2 for feature extraction, and the audio modality utilizing MMV for feature extraction. These features are then fed into a multi-scale Actionformer.}
\label{fig:short}
\end{figure}

\section{Method}
\subsection{Overall Architecture}
In our work, we use Actionformer as our base model, which combines multi-scale feature representations with local self-attention. It employs a lightweight decoder to classify each moment and estimate the corresponding sound boundaries. Additionally, we incorporate multimodal features as input, where the visual features are obtained from the self-supervised pre-trained network VideoMAE V2 \cite{DBLP:conf/cvpr/WangHZTHWWQ23,yang2019adaptive}, and the audio features from the MMV audio embeddings pre-trained on AudioSet \cite{DBLP:conf/nips/AlayracRSARFSDZ20}. By combining the video and audio features, we achieve multimodal input.

\subsection{Multimodal Feature Extraction}
For the visual modality, in order to obtain better visual representations, we use VideoMAE V2 to extract visual features. This model employs a dual-mask strategy for efficient self-supervised pre-training and has achieved state-of-the-art results on various downstream video tasks.

For the audio features, the MMV audio feature from [2] is pre-trained on AudioSet. The representation is as follows:
\vspace{-5pt}
\begin{equation}
\begin{split}
    Z_{v} =\left \{ z^{1} _{v}, z^{2} _{v}, ...,z^{L} _{v}  \right \}, Z_{v} = g\left ( X_{v}  \right ) \\
    Z_{a} =\left \{ z^{1} _{a}, z^{2} _{a}, ...,z^{L} _{a}  \right \}, Z_{a} = f\left ( X_{a}  \right )\\
    Z_{0} = Z_{v} \oplus Z_{a}
\end{split}
\end{equation}

Where $Z_{v}$ represents the visual features of the video, $g\left ( \cdot  \right ) $ denotes the visual feature extraction network. $Z_{a}$ represents the audio features of the video, and $f\left ( \cdot  \right ) $ denotes the audio feature extraction network. The $\oplus$ symbol indicates feature concatenation.

\subsection{Multi-scale Feature Sound Localization}

We use $Z_{0}$ as the input to the Transformer network, and its core design lies in the calculation process of self-attention. Actionformer adopts local self-attention calculation to avoid the decrease in performance caused by long attention windows. In order to better capture multi-scale features, it incorporates downsampling operations into the network structure. The downsampling operation uses 1D convolution and employs a downsampling stride of 2x for each layer.  This is given by
\vspace{-5pt}

\begin{equation}
\begin{split}
    \bar{Z^{\ell }} = \alpha ^{\ell } MSA\left ( LN\left ( Z^{\ell -1} \right )  \right ) , \\\widehat{Z} ^{\ell} = \bar{\alpha ^{\ell }} MLP\left ( LN\left ( \bar{Z^{\ell}} \right )  \right ) + \bar{Z^{\ell }}\\
    Z^{\ell} =\downarrow \left ( \widehat{Z} ^{\ell}  \right ) , \ell=1,...L,
\end{split}
\end{equation}

Where $Z^{\ell - 1}, \bar{Z^{\ell }}, \widehat{Z} ^{\ell} \in \mathbb{R} ^{T^{l-1}\times D}$ and $Z^{\ell} \in \mathbb{R} ^{T^{l}\times D}$. $T^{\ell -1} /T^{\ell}$ is the downsampling ratio. $\alpha ^{\ell }$ and $\bar{\alpha ^{\ell }}$ are learnable per-channel scaling factors as in \cite{DBLP:conf/iccv/TouvronCSSJ21}.

The decoder is a lightweight convolutional network with a classification and a regression head. We keep only 17 sound classes. And we trained the transformer blocks and the classification and regression heads to accommodate the number of classes included in our dataset. The model outputs $\left ( p\left ( a_{t} \right ), d^{s}_t, d^{e}_t    \right )$ for every moment $t$, including
the probability of action categories $p\left ( a_{t} \right )$ and the distances to action boundaries $\left ( d^{s}_t, d^{e}_t \right ) $. Our loss function, again following minimalist design, only has two terms:(1) $\mathcal{L}_{cls}$  a focal loss for C way binary classification; and (2) $\mathcal{L}_{reg}$ a DIoU loss for distance regression. The loss is as follows:
\vspace{-3pt}
\begin{align}
\mathcal{L} =\sum_{t}^{} \left ( \mathcal{L}_{cls} + \lambda_{reg} \mathbb {I}_{c_t}  \mathcal{L}_{reg} \right )/T_{+}  
\end{align}

\vspace{-5pt}

\section{Experiment}
\textbf{Dataset.} This dataset is provided by the official competition organizers and comprises approximately 11,433 videos. And the number of annotations is 137,128. The train video number is 2149, the validation video number is 5359 and the test video number is 3238.

\textbf{Metric.} The evaluation metric for this Task is the mean Average Precision (mAP).

\textbf{Implementation Detail.} In our study, our experimental setup follows VideoMAE V2 and Actionformer. The feature dimension for the visual modality is 1408, and for the audio modality, it is 256. Therefore, the input feature dimension to Actionformer is 1564. A window size of 11 was used for local self-attention. The number of Transformer blocks is 9. We only use a single NVIDIA RTX 3090 to train the model, we set the initial learning rate as 1e-4, the batch size as 2, and the epoch as 35.

\begin{table}
\centering
\label{tab: map}
\begin{tabular}{@{\hspace{0.08cm}}c@{\hspace{0.08cm}}c@{\hspace{0.08cm}}c@{\hspace{0.08cm}}c@{\hspace{0.08cm}}c@{\hspace{0.08cm}}c@{\hspace{0.08cm}}c@{\hspace{0.08cm}}}
\toprule
Method & @0.1 & @0.2 & @0.3 & @0.4 & @0.5 & Avg\\
\hline
Baseline(video+audio) & 18.8 & 17.6 & 15.9 & 13.9 & 11.3 &  15.5 \\
Ours(video) & 41.5 & 37.2 & 34.1 & 29.7 & 20.1 &  30.5 \\
Ours(audio) & 16.2 & 13.5 &  10.8 &  8.4 & 5.8 &  10.9 \\
Ours(video+audio) & 41.7 & 38.2 & 34.6 & 30.3 & 20.7 &  33.1 \\
\toprule
\end{tabular}
\caption{ Mean average precision (mAP) for temporal sound localization task using ActionFormer as baseline and our method. IoU for 0.1-0.5 are averaged as in \cite{DBLP:journals/ijcv/DamenDFFKMMMPPW22}.}
\end{table}

\textbf{Result.} Table 1 shows the mAP score performance. Baseline (video+audio) is the method provided by the official source, using TSN-based visual features and MMV audio features. When we replace it with features extracted by VideoMAE V2, there is a significant improvement in mAP. From the experiments, it can be observed that the audio modality is a weak modality, and the improvement after feature fusion is small.

\section{Conclusion}
This report summarized our solution for the Sound Localisation Task in the ICCV 1st Perception Test Challenge 2023. Our approach combines advanced visual features and audio features, which are then input into a multi-scale Transformer for training, resulting in effective results. We achieved the 2nd performance on this track.

\bibliographystyle{plain}
\bibliography{main}

\end{document}